\documentclass[sigconf]{acmart}
\usepackage{booktabs}
\usepackage{multirow}
\usepackage{graphicx}
\usepackage[most]{tcolorbox}
\usepackage{listings}

\usepackage{tikz}
\usepackage[table]{xcolor}
\usepackage{booktabs}
\usepackage[rightmargin=0.3pt,leftmargin=0.3pt]{quoting}

\newtcolorbox{surveybox}{
  enhanced,
  arc=1mm,
  colback=white,
  colframe=black,
  boxrule=0.7pt,
  width=\textwidth, 
  left=4mm,
  right=4mm,
  top=3mm,
  bottom=3mm,
  before skip=10pt,
  after skip=10pt,
}

\definecolor{NotAtAll}{gray}{0.9}
\definecolor{Slightly}{gray}{0.75}
\definecolor{Moderately}{gray}{0.55} 
\definecolor{Very}{gray}{0.35}
\definecolor{Extremely}{gray}{0.2}

\newcommand{\progressbar}[7]{%
  \begin{tikzpicture}[baseline={(current bounding box.center)}]
    \node[font=\small] at (-0.6, 0.2) {#6};
    \fill[NotAtAll]   (0,0)                    rectangle (#1,0.25);
    \fill[Slightly]   (#1,0)                   rectangle (#1+#2,0.25);
    \fill[Moderately] (#1+#2,0)                rectangle (#1+#2+#3,0.25);
    \fill[Very]       (#1+#2+#3,0)             rectangle (#1+#2+#3+#4,0.25);
    \fill[Extremely]  (#1+#2+#3+#4,0)          rectangle (#1+#2+#3+#4+#5,0.25);
    \node[font=\small] at (#1 + #2 + #3 + #4 + #5 + 0.5, 0.2) {#7};
  \end{tikzpicture}%
}

\setcopyright{acmlicensed}
\copyrightyear{2025}
\acmYear{2018}
\acmDOI{XXXX}
\acmConference[acronym 'XX]{title}{date}{NY}

\acmISBN{978}

\begin{document}

\title{Beyond Code: Empirical Insights into How Team Dynamics Influence OSS Project Selection}

\author{Shashiwadana Nirmani}
\affiliation{%
  \institution{Deakin University}
  \country{Australia}
}
\email{s.dona@research.deakin.edu.au}

\author{Hourieh Khalajzadeh}
\affiliation{%
  \institution{Deakin University}
  \country{Australia}
}
\email{hourieh.khalajzadeh@deakin.edu.au}

\author{Mojtaba Shahin}
\affiliation{%
  \institution{RMIT University}
  \country{Australia}
}
\email{mojtaba.shahin@rmit.edu.au}

\author{Xiao Liu}
\affiliation{%
  \institution{Deakin University}
  \country{Australia}
}
\email{xiao.liu@deakin.edu.au}

\begin{abstract}
Open-source software (OSS) development relies on effective collaboration among distributed contributors. Yet, current OSS project recommendation systems primarily emphasize technical attributes, overlooking the collaboration and community aspects that influence contributors’ decisions to join and remain in projects. This study investigates how team dynamics within OSS communities influence project selection and how these preferences vary across contributors’ motivations. We conducted an online survey with 198 OSS practitioners, combining quantitative and qualitative analyses to capture contributors’ perceptions of team dynamics. The results reveal that communication-related team dynamics such as responsiveness, tone, and clarity of replies are consistently prioritized across practitioners. However, the relative importance of these team dynamics differs according to contributors’ motivations. For instance, practitioners motivated by gaining reputation or networking preferred inclusive project communities that encouraged diverse participation. These findings highlight that understanding how team dynamics align with contributors’ motivations provides valuable insights into practitioners’ project selection behaviour. Those insights can inform the design of future human-aware project recommendation systems that better account for social collaboration quality and motivational fit.
\end{abstract}


\maketitle

\section{Introduction}
Open-source software (OSS) development has become one of the most successful examples of large-scale, distributed collaboration in software engineering \cite{linaaker2025public,yue2025discovering}. However, despite the technical maturity of many projects, maintaining effective collaboration remains a challenge \cite{steinmacher2019overcoming}. Contributors often encounter barriers such as unconstructive feedback, lack of responsiveness, and unclear team processes. These factors significantly affect whether individuals choose to join or stay active in a project \cite{steinmacher2019overcoming,iaffaldano2019developers,miller2022did}. These social challenges can undermine productivity and community health even when the technical foundations of a project are strong.

Existing literature has identified several areas of collaboration, such as social and emotional interactions, communication patterns, and coordination mechanisms, which play a major role in the functioning and sustainability of OSS projects \cite{qiu2019signals,guizani2021long,steinmacher2019overcoming,gousios2016work}.  For instance, emotionally intelligent communication supports conflict resolution, trust, and satisfaction, while toxic interactions or dismissive responses can discourage participation \cite{miller2022did}. OSS projects differ from co-located teams in that contributors rely on asynchronous communication and limited social context, making interpersonal signals such as tone, responsiveness, and inclusivity vital indicators of team health \cite{qiu2019signals}. In this study, we refer to the behavioral and interpersonal aspects of collaboration that shape how contributors interact, communicate, and coordinate within a project, as \textbf{team dynamics} \cite{guizani2021long,steinmacher2019overcoming,qiu2019signals}.


The current OSS project recommendation systems continue to focus predominantly on technical attributes such as programming language, star count, tags \cite{nirmani2025systematic,phatangare2024codecompass,shen2024multi,abbasi2021relevant}. Such approaches overlook the social texture of projects, including the interactions, responsiveness, and support systems that influence practitioners’ motivation to engage. Prior work has shown that contributors evaluate cues like issue discussions, documentation tone, and responsiveness before deciding to contribute \cite{qiu2019signals}. Yet, a systematic understanding of which team dynamics matter most to practitioners remains lacking.

Motivation is one of the key human factors influencing participation in OSS communities \cite{nirmani2025systematic}. Contributors engage in projects for diverse reasons such as learning new skills, helping others, and gaining enjoyment \cite{guizani2021long, wu2007empirical, roberts2006understanding}. These underlying motivations can shape how they perceive and value different aspects of collaboration. For example, contributors motivated by learning may prioritize teams that provide timely feedback and mentorship, while those motivated by enjoyment or helping may prefer inclusive and supportive environments. Understanding how motivations influence preferences for team dynamics provides deeper insight into why practitioners choose particular projects and how social factors contribute to sustained engagement.

To address these gaps, we conducted an online survey with 198 OSS practitioners to investigate \textit{what team dynamics (e.g., having active members contributing regularly, gender diversity within the team, tone and manner of communication) influence their project selection (RQ 1)} and \textit{how these preferences vary across motivational dimensions (RQ 2)}. The survey combined both quantitative and qualitative analyses to capture practitioners’ perceptions. 

The results revealed that communication-related team dynamics were consistently prioritized by the participants (RQ 1.1). Specifically, practitioners highly valued the \textit{consistency of receiving responses to questions}, \textit{clarity and simplicity of replies}, and the \textit{tone and manner of communication}. They also emphasised responsiveness indicators such as the \textit{time taken to review pull requests (PR)} and \textit{typical response time to questions or inquiries.}
The open-ended responses (RQ 1.2) further highlighted additional desirable team dynamics such as \textit{willingness to collaborate}, \textit{onboarding resources and assistance}, and \textit{inclusive and transparent decision-making}, reflecting a preference for inclusive and supportive collaboration environments. However, the relative importance of these team dynamics varied across motivational categories (RQ2). For instance, contributors with a high \textit{networking and socializing} motivation valued \textit{diverse and inclusive community}; while participants with high \textit{enjoyment} motivation prioritized \textit{beginner-friendly, structured} and \textit{active collaboration} environments.

By uncovering the importance of different team dynamics that contributors consider and the influence of motivational factors on team dynamics, our findings have broader implications for OSS practitioners and researchers. First, the strong emphasis on responsiveness and respectful communication suggests opportunities for human–AI co-maintenance of OSS projects that can support contributors through emotion-aware AI assistants. Second, the link between inclusive communication and perceived diversity highlights the need to design communities that foster belonging and fairness across demographic and regional boundaries. Third, the preference for transparent and collaborative decision-making reinforces the importance of participatory governance models that distribute authority and promote fairness. Finally, understanding how motivations interact with team dynamics enables motivation-aware recommender systems that go beyond technical fit to support sustainable engagement and retention.

The remainder of this paper is organized as follows. Section \ref{related_work} outlines the Related Work, Section \ref{Method} presents the Methodology, Section \ref{results} presents the Results, Section \ref{discussion} outlines the Discussion, and Section \ref{threats} discusses the Threats to Validity. Finally, Section \ref{conclusion} concludes the paper and outlines some potential future work.

\section{Related Work} \label{related_work}
\subsection{Team Dynamics in OSS} \label{team_dynamics_lr}
Numerous studies in OSS highlight that social collaboration-related factors strongly shape OSS contributor participation and retention. Steinmacher et al. \cite{steinmacher2014barriers} identified \textit{delayed or missing responses} as one of the most common onboarding barriers, while \textit{unhelpful, complex or unclear replies} discourage further contributions \cite{steinmacher2019overcoming}. Moreover, \textit{not accepting newcomers' contributions or revising them without proper explanation}, \textit{ignoring PRs and leaving them open indefinitely}, and \textit{delay in resolving issues reported by users} are recorded as unattractive work practices in OSS communities.  \cite{guizani2021long, gousios2016work, steinmacher2015social, jarczyk2014github}. Similarly, research on OSS communication threads found that \textit{hostile or impolite tones} and \textit{not having active discussion threads} are reported as major reasons for contributors to leave OSS projects \cite{ehsani2025analyzing,guizani2021long,steinmacher2014barriers}. However, another research reported that PRs with longer discussions are less likely to get accepted \cite{tsay2014influence}. 

Moreover, \textit{having focused team members} rather than a few popular contributors has been shown to support smoother project maintenance and task coordination \cite{jarczyk2014github}. Gousios and Bacchelli \cite{gousios2016work} reported that \textit{inactive teams }create frustration and discourage sustained participation.
\textit{Not actively maintaining the project} results in projects appearing abandoned and discourages continued involvement \cite{prana2021including}.
Turzo et al. \cite{turzo2025first} showed that repositories using \textit{descriptive tags} such as “help wanted” and “good first issue (GFI)” provide effective entry points for newcomers and increase the likelihood of long-term engagement. Tan et al. \cite{tan2020first} revealed that while the \textit{GFI mechanism} helps attract newcomers, its effectiveness depends on consistent labelling, clear task descriptions, and active maintainer support to prevent discouragement among first-time contributors.

Research on \textit{diversity and inclusion} reveals persistent disparities. Terrell et al. \cite{terrell2017gender} demonstrated gender bias in PR acceptance. Their results revealed that women’s contributions were accepted more frequently than men’s overall; however, when contributors were outsiders and their gender was identifiable, men’s pull requests received higher acceptance rates \cite{terrell2017gender}. Similarly, Prana et al. \cite{prana2021including} conducted a large-scale analysis of GitHub repositories across multiple \textit{geographic regions} and found that gender and regional diversity remain low globally. Developers from the same region were often found to have higher pull-request acceptance rates, reflecting implicit regional familiarity and shared linguistic or cultural norms \cite{prana2021including}. The authors emphasized that language barriers, limited access to resources, and differing regional development opportunities contribute to uneven participation levels. To address these disparities, they recommended several strategies: fostering friendlier and safer communities through enforceable codes of conduct; establishing mentorship and role-model visibility programs to highlight developers from underrepresented regions; and enhancing tooling support, such as translation and automated documentation localization, to reduce linguistic barriers and broaden participation \cite{prana2021including}. 

\subsection{Motivations in OSS}

Several studies have examined the various motivational factors that encourage individuals to participate in OSS projects. Ye and Kishida \cite{ye2003toward} highlighted learning as a major intrinsic motivator, suggesting that modular system designs with independent tasks of increasing complexity can help newcomers begin with simpler tasks and progressively take on more challenging ones. Wu et al. \cite{wu2007empirical} reported that OSS contributors’ sustained participation is influenced by both altruistic motives and financial rewards. In contrast, Hartman \cite{hartman2011intrinsic} observed that monetary compensation might undermine autonomy and intrinsic motivation by introducing social pressures. Bitzer et al. \cite{bitzer2007intrinsic} focused on intrinsic motivations, emphasizing the influence of young and skilled programmers who contribute out of enjoyment and alignment with the OSS gift culture.

Leadership as a motivational factor has been examined in several studies \cite{li2012leadership, wynn2004leadership} which found that transformational leadership enhances intrinsic motivation, participative leadership increases engagement, and achievement-oriented leadership fosters a sense of appreciation.
Von et al. \cite{von2012carrots} identified five key factors influencing motivation: licensing restrictions, community sponsorship, governance structures, rewards, and social or technical exposure. Gerosa et al. \cite{gerosa2021shifting} investigated a range of motivational dimensions, including ideology, altruism, enjoyment, kinship, reputation, reciprocity, learning, personal use, career advancement, and pay. They further explored how motivations evolve as contributors gain experience.

\subsection{OSS Project Recommendation Systems and Team Dynamics}

The number of OSS project recommendation systems has increased in recent years, with most focusing on a project’s technical features \cite{nirmani2025systematic}. For instance, systems often analyze contributors’ past actions, such as commenting, forking, or starring, to generate recommendations \cite{liao2023graph,sayce2022recommendation,shen2024multi}. They also consider contributors’ skills derived from previous work or public profiles \cite{phatangare2024codecompass,shen2024multi}. Similarly, project-related attributes like popularity, domain, title, description, and README content are commonly used to identify suitable projects and match them with potential contributors \cite{ford2022reboc,liao2023graph,sayce2022recommendation}. However, none of the existing studies have considered the team dynamics of OSS communities in their project recommendation systems. 

Several studies have begun to incorporate social and behavioural preferences, but they do so primarily at the individual user level rather than the project community level. For instance, \cite{li2016task} proposed a social influence–based recommendation approach that constructs a developer social network from behaviors such as browsing, bidding, and completing tasks, and then measures social influence degrees between developers. Their method demonstrated improved task matching for both active and inactive developers by leveraging the interactions of socially connected peers, effectively addressing the cold-start problem. They noted that the time it takes for a user to respond to the recommendation will be considered for future improvements to their system. 

Similarly, Santos et al. \cite{santos2023tell} analyzed developers’ communication networks to identify those who occupy central roles in issue discussions, showing that such engagement patterns can reveal domain expertise.
They modelled communication networks using social metrics, including the number of comments and commenters, to capture developers’ roles in conversations. These results highlight that patterns of dialogue and participant connectedness can signal both the technical relevance of an issue and the skill domains of involved developers, reinforcing the value of communication networks for improving task or issue recommendations.

Collectively, these works demonstrate that social signals can enhance project matching, but they primarily focus on modelling the contributor rather than the project’s community. As summarized in our review, no existing recommender system incorporates these team-level dynamics of OSS communities. This motivates our study’s focus on characterizing team dynamics to be used in project recommendation systems.

\section{Methodology} \label{Method}
In this section, we present our Research Questions (RQs) and outline our methodology. To address our RQs, we conducted an online survey targeting practitioners with experience in OSS contributions. The materials necessary to replicate this work including the complete survey questionnaire are available in \cite{anonymous_2025_beyondcode}.

\subsection{Research Questions}

\begin{tcolorbox}[arc=1mm,width=1.0\columnwidth,
                  top=1mm,left=1mm,  right=1mm, bottom=1mm,
                  boxrule=.75pt]
 \textbf{RQ 1: What team dynamics influence OSS practitioners’ decisions to join projects?}
\end{tcolorbox}

This RQ aims to identify what team dynamics OSS practitioners consider when selecting an OSS project to contribute to. The RQ is studied under two sub-questions, as described below.

\textit{\textbf{RQ 1.1}: How do practitioners rate the importance of team dynamics when selecting OSS projects?} 
This sub-question examines the relative importance of predefined team dynamics such as responsiveness, communication tone, and team diversity to identify which factors consistently drive participation.
\vspace{0.5em}

\textit{\textbf{RQ 1.2}: What additional team dynamics do practitioners identify as important beyond those analyzed in RQ 1.1?}
This sub-question complements RQ1.1 by capturing additional team dynamics mentioned by practitioners, offering an open perspective on overlooked or emerging team behaviors that shape their decisions.

\vspace{0.5em}

\begin{tcolorbox}[arc=1mm,width=1.0\columnwidth,
                  top=1mm,left=1mm,  right=1mm, bottom=1mm,
                  boxrule=.75pt]
 \textbf{RQ 2: How do OSS practitioners’ motivations influence team dynamics when deciding to join OSS projects?}
\end{tcolorbox}


Existing research has demonstrated that OSS project maintainers can strategically manage their projects to sustain the engagement of highly skilled contributors. For instance, Smirnova et al. \cite{smirnova2022makes} found that project owners are able to motivate their most talented volunteer contributors when they show generosity in accepting external contributions and provide prompt, constructive feedback. These findings highlight how team dynamics, particularly responsiveness and openness in communication, play a crucial role in reinforcing contributors’ motivation to continue participating in OSS projects. Another study found that newcomers closely observe how friendly and supportive the project communication is, and negative or unfriendly interactions can quickly demotivate them from continuing to contribute \cite{miller2022did}. Psychological safety is linked to motivating both short-term and long-term participation \cite{sesari2025safe}. 

Hence, this RQ aims to provide a comprehensive analysis of the influence of practitioners' motivations on their preferences for team dynamics. By linking motivation to perceived team dynamics, this RQ provides deeper insight into how to retain and sustain contributors by providing personalized project recommendations aligned with their motivational and collaboration preferences.

\subsection{Survey Design}
We designed an online survey that included both qualitative and quantitative questions, following the guidelines of Kitchenham and Pfleeger \cite{kitchenham2008personal}. The survey was fully anonymous, ensuring no participant could be identified, and it was approved by the institutional ethics committee. Participants were informed about the study’s purpose and details in the survey preamble. The estimated completion time was 8-10 minutes.

Based on the literature, the survey was organised into several parts: screening questions (Section \ref{screening_process}), team dynamics, motivations for joining OSS, demographics and two open-ended questions to provide which additional team dynamics they would consider when joining OSS and any additional comments they have. Participants could choose “Prefer not to say” for gender and region, and all questions except the open-ended questions were mandatory. In total, the survey contained 41 questions: 6 multiple-choice screening questions, 19 team dynamics questions (3 multiple-choice and 16 Likert-scale), 7 Likert-scale motivation questions, 7 demographic questions, and 2 open-ended questions. A pilot survey was first conducted with five practitioners recruited through personal contacts. They provided feedback on minor unclear wordings, which were refined before launching the final version.

\subsubsection{Survey Questionnaire}
As mentioned in Related work \ref{related_work} - Section \ref{team_dynamics_lr}, prior research has explored different team dynamics in OSS communities \cite{guizani2021long, steinmacher2019overcoming,steinmacher2015social,bosu2019diversity}. In this study, a set of team dynamics (e.g., consistency of receiving responses to questions, gender diversity, tone and manner of communication) was selected based on recurring themes identified across prior OSS literature, particularly those shown to influence contributor participation and collaboration quality, such as responsiveness, communication tone, inclusiveness, and activity level. These factors were refined and consolidated into measurable survey items to ensure broad coverage of key social and coordination aspects relevant to OSS teams. A summary of our survey questionnaire is mentioned in Figure \ref{questions}.

In this study, we examined the influence of seven highly cited motivations on team dynamics, which are referred to by the following abbreviated terms for clarity and consistency throughout the paper. \textit{Learning or gaining new skills (Learning)} \cite{ke2009motivations, wu2007empirical, roberts2006understanding, hertel2003motivation,lakhani2003hackers}, where contributors aim to enhance their knowledge and abilities; \textit{Helping the programmer community (Helping)} \cite{ke2009motivations, wu2007empirical,lakhani2003hackers}, reflecting a desire to support and give back to the community; \textit{Enjoyment (Enjoyment)} \cite{ke2009motivations, roberts2006understanding,  lakhani2003hackers}, where contributors participate because they find the activity engaging and enjoyable; \textit{Career advancement (Career)} \cite{ke2009motivations, wu2007empirical, roberts2006understanding, hertel2003motivation, lakhani2003hackers}, viewing OSS participation as a way to improve employability or professional growth; \textit{Networking and socializing (Networking)} \cite{von2012carrots, ke2009motivations, hertel2003motivation}, focusing on building professional and social connections; \textit{Gaining a reputation (Reputation) }\cite{ke2009motivations, roberts2006understanding,  hertel2003motivation, lakhani2003hackers}, which enhances credibility within the community; and \textit{Financial incentives (Incentives)} \cite{von2012carrots, roberts2006understanding, lakhani2003hackers}, where contributors are motivated by monetary or material rewards.
\begin{figure*}[t]
\centering
\begin{surveybox}

\textbf{QUESTION 1:} How important is it for a project to have active members contributing regularly?
\\[0.25em]
\textbf{OPTIONS}: \texttt{Not at all important | Slightly important | Moderately important | Very important | Extremely important  } 

\textbf{QUESTION 2:} How does team diversity influence your decision to contribute to OSS? (Scale: Not at all important - Extremely important) \\[0.25em]
\textbf{OPTIONS}: \texttt{Gender diversity | The presence of contributors from various countries | The presence of contributors from your own country } 

\textbf{QUESTION 3:} How does communication practices within project team influence your decision to contribute? (Scale: Not at all important - Extremely important) \\[0.25em]
\textbf{OPTIONS}: \texttt{The consistency of receiving responses to questions | The typical response time to questions or inquiries| The clarity and simplicity of replies to questions| The tone and manner of communication (e.g., respectful, inclusive, or welcoming) } 

\textbf{QUESTION 4:} How does team practices influence your decision to contribute? (Scale: None at all - A great deal) \\[0.25em]
\textbf{OPTIONS}: \texttt{The frequency of merging pull requests from new contributors | Whether pull requests generally receive a clear response (either accepted or rejected) | The typical time taken to review or respond to pull requests | The time taken to resolve reported issues | The availability of beginner-friendly issues (e.g., labeled as "Good First Issue") | The use of descriptive labels on issues (e.g., "wontfix", "help wanted", "bug") | The frequency of recent code commits | The recency of project releases | The level of activity in issue or pull request discussions} 

\textbf{QUESTION 5:} Please rate how much each of the following motivations influences your decision to contribute to OSS projects, using the scale: Not at all, Slightly, Moderately, Very, Extremely \\[0.25em]
\textbf{OPTIONS}: \texttt{Learning or gaining new skills| Helping the programmer community | Enjoyment | Career advancement | Networking and socializing | Gaining a reputation | Financial incentives} 

\textbf{QUESTION 6:} Beyond the given factors like contributors' reputation, welcoming communication style, etc., are there any other team dynamics you would consider when joining an OSS project? Describe how these factors would influence your decision.

\end{surveybox}
\caption{Summary of Survey Questionnaire}
\label{questions}
\end{figure*}

\subsubsection{Screening Process} \label{screening_process} 

    Participants were first asked two basic screening questions: (1) “Do you have experience in OSS development?” and (2) “Which of the following OSS platforms or repositories are you familiar with?”. Those who successfully passed these initial checks were then presented with four technical screening questions adapted from Danilova et al.  \cite{danilova2021you}. To prevent overwhelming the participant while maintaining accuracy, we selected Q1 (which includes two sub-questions), Q15, and Q16. These questions tested familiarity with lesser-known programming languages, understanding of code fragments (e.g., identifying a function parameter), and identifying the correct return value of pseudocode. Participants needed to pass all screening questions to proceed.

\subsection{Recruitment}

Survey participants were recruited through multiple platforms to achieve a diverse sample. Invitations were distributed via Prolific \cite{Prolific} (an online recruitment platform commonly used in software engineering research \cite{reid2022software, russo2024navigating, tahaei2023stuck}) and through social media, such as LinkedIn \cite{LinkedIn}. We enabled the “software industry” domain filter on Prolific to better target relevant participants. Participants who failed any of the screening questions (Section \ref{screening_process}) were deemed ineligible. We sent 650 invitations, only 170 (26\%) were able to pass the screening questions and complete the survey. Prolific participants received monetary compensation for their time. Additionally, we received 33 responses from social media and personal contacts. However, 5 incomplete responses were excluded from the analysis. Hence, the final analysis included 198 responses.

\subsection{Analysis}
For data analysis, we used a combination of statistical and thematic analysis methods. The quantitative data were examined using standard statistical analysis techniques, following the guidelines recommended by Kitchenham and Pfleeger for survey analysis \cite{kitchenham2008personal}.

To identify statistically significant differences (RQ2), we applied the Kruskal–Wallis test to compare team dynamic preferences and motivation preferences \cite{mckight2010kruskal}. The Kruskal–Wallis test is a non-parametric method that extends the Mann-Whitney U test \cite{mcknight2010mann} to compare more than two independent groups without assuming a normal distribution. It evaluates whether three or more sample groups originate from the same distribution on a variable of interest. This test is commonly used in software engineering research \cite{aldndni2024understanding,meem2024exploring,ouyang2025empirical}. A p-value of $\leq 0.05$ was considered statistically significant, indicating a statistically significant difference in the mean ranks across the groups \cite{aldndni2024understanding}. Additionally, we performed a post-hoc Bonferroni correction on P values to adjust for multiple comparisons and minimize false-positive results. \cite{weisstein2004bonferroni}.

We analyzed the qualitative data obtained from the open-ended question using thematic analysis, which followed the standard phases of familiarizing with the data, generating initial codes, identifying potential themes, reviewing them, and finalizing theme definitions \cite{braun2006using}. The first author initially read all responses to gain an overall understanding and then divided them into 190 meaningful segments for coding. These segments were categorized into preliminary themes. The second and third authors independently reviewed the codes and themes noting any inconsistencies or concerns. Differences were discussed collaboratively until full consensus was reached, resulting in a final agreed-upon set of themes. During this process, 46 responses were removed due to insufficient or unclear content. To assess coding reliability, Cohen’s Kappa was computed, yielding a score of 0.91 \cite{tahaei2023stuck}, indicating a high level of inter-annotator agreement and confirming the robustness of the qualitative analysis. 

\section{Results} \label{results}


Survey participants had diverse demographic backgrounds. Age-wise, they were in 18-24 (n=38), 25-34 (n=102), and 35 and over (n=58) categories. Regarding gender, 44 participants identified as female, 150 as male, 2 as other, and 2 preferred not to say. Regionally, Africa (n=20), the Americas (n=27), Asia (n=27), Europe (n=116), and Oceania (n=5), prefer not to say (n=3). In terms of OSS contribution experience, 64 participants had less than 1 year of experience, 97 had 1-3 years, 28 had 4-6 years, and 9 had 7 years or more. 

\subsection{RQ 1: What team dynamics influence OSS practitioners’ decisions to join projects?} \label{team_dynamics_1.1}
\subsubsection{RQ 1.1: How do practitioners rate the importance of team dynamics when selecting OSS projects?} \label{team_dynamics_1_1_1}

Table \ref{rq1_analysis} highlights how practitioners rate the importance of team dynamics when considering whether to join OSS projects. In the description below, the reported percentages refer to participants who rated each factor as “Very” or “Extremely” important. OSS practitioners place the greatest importance on communication-related dynamics, particularly \textit{consistency of receiving responses to questions} (76\%), \textit{clarity and simplicity of replies to questions} (76\%), and \textit{tone and manner of communication} (e.g., welcoming and respectful) (73\%). 
\textit{Having active members contributing regularly} (70\%) and responsiveness-related factors, such as \textit{whether PRs generally receive a clear response} (either accepted or rejected) (69\%), \textit{typical response time to inquiries} (58\%), \textit{time taken to resolve reported issues} (57\%), \textit{level of activity in issue or PR discussions} (57\%) and \textit{time taken to review a PR} (56\%), were also seen as highly influential. 

Moreover, the participants emphasized the importance of fast and regular replies in the open-ended responses. Similarly, \textit{the use of descriptive labels} such as ``Help Wanted'' (53\%) and \textit{availability of beginner-friendly issues} such as ``Good First Issues" (52\%) were also valued by them. Additionally, \textit{the frequency of recent code commits} (46\%), \textit{the recency of project releases} (46\%) and \textit{the frequency of merging PRs from new contributors} (38\%) were also deemed highly important. In contrast, diversity-related aspects such as \textit{gender diversity} (17\%), \textit{having contributors from their own country} (15\%) or \textit{having contributors from various countries (20\%)} were generally rated lower. However, the broader theme of fostering diversity and inclusion reappeared in the open-ended responses. Overall, these findings suggest that practitioners prioritize fast, constructive, and reliable interactions, while also recognizing the complementary role of diversity and inclusion in shaping effective team dynamics.

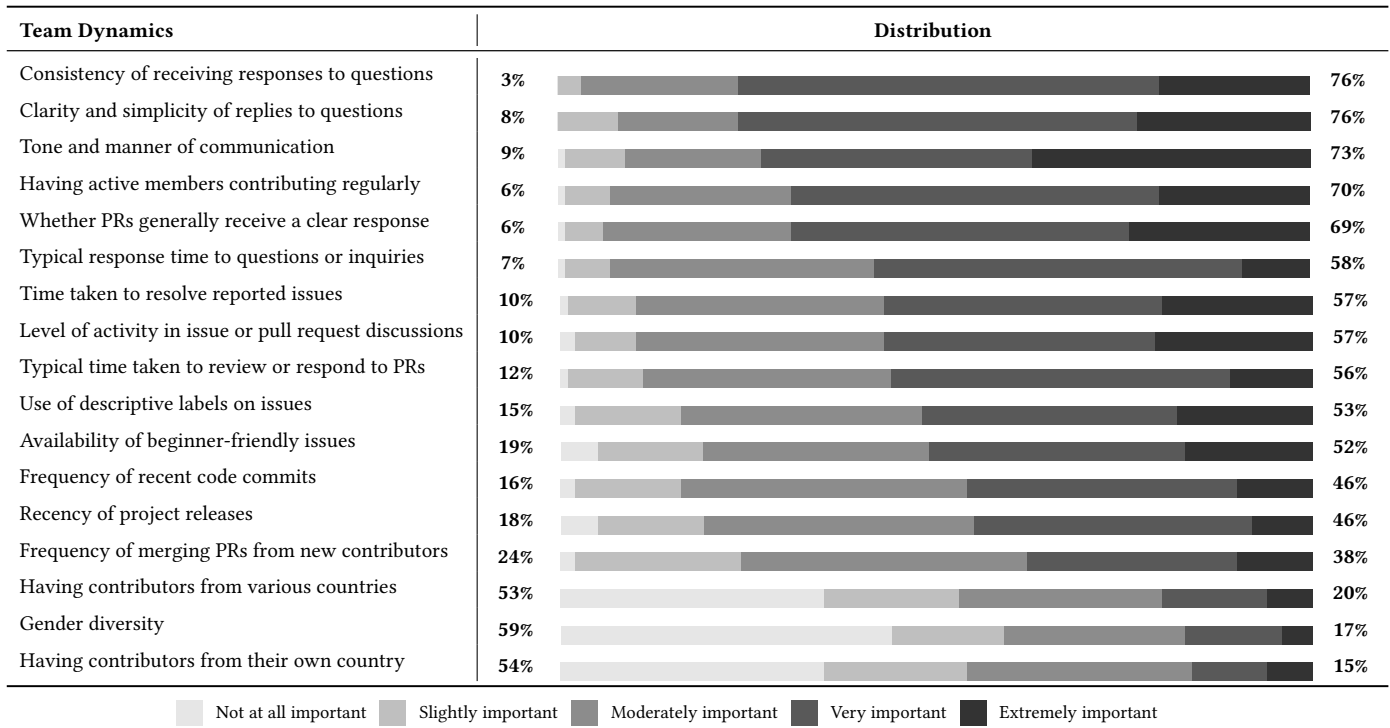
\begin{table*}[]
    \centering
    \small
    \renewcommand{\arraystretch}{1.2} 
    \setlength{\tabcolsep}{5pt} 

     \caption{Distribution of Team Dynamics Preferences. Percentages in bold on the chart (N\%) represent the percentage of the distribution that reported “Not at all”/“Slightly” (left) and “Very”/“Extremely” (right) important.}
     \label{rq1_analysis}
     \begin{tabular}{l | c}
        \toprule
        \textbf{Team Dynamics} & \textbf{Distribution} \\
        \midrule
        Consistency of receiving responses to questions     & \progressbar{0}{0.3}{2.1}{5.6}{2}{\textbf{3\%}}{\textbf{76\%}} \\ 
        Clarity and simplicity of replies to questions & \progressbar{0}{0.8}{1.6}{5.3}{2.3}{\textbf{8\%}}{\textbf{76\%}} \\
        Tone and manner of communication  & \progressbar{0.1}{0.8}{1.8}{3.6}{3.7}{\textbf{9\%}}{\textbf{73\%}} \\
        Having active members contributing regularly    & \progressbar{0.1}{0.6}{2.4}{4.9}{2} {\textbf{6\%}}{\textbf{70\%}} \\
        Whether PRs generally receive a clear response  & \progressbar{0.1}{0.5}{2.5}{4.5}{2.4}{\textbf{6\%}}{\textbf{69\%}} \\
        Typical response time to questions or inquiries   & \progressbar{0.1}{0.6}{3.5}{4.9}{0.9}{\textbf{7\%}}{\textbf{58\%}} \\
        Time taken to resolve reported issues  & \progressbar{0.1}{0.9}{3.3}{3.7}{2}{\textbf{10\%}}{\textbf{57\%}} \\
        Level of activity in issue or pull request discussions  & \progressbar{0.2}{0.8}{3.3}{3.6}{2.1}{\textbf{10\%}}{\textbf{57\%}} \\
        Typical time taken to review or respond to PRs & \progressbar{0.1}{1}{3.3}{4.5}{1.1}{\textbf{12\%}}{\textbf{56\%}} \\ 
        Use of descriptive labels on issues  & \progressbar{0.2}{1.4}{3.2}{3.4}{1.8}{\textbf{15\%}}{\textbf{53\%}} \\
        Availability of beginner-friendly issues & \progressbar{0.5}{1.4}{3}{3.4}{1.7}{\textbf{19\%}}{\textbf{52\%}} \\
        Frequency of recent code commits  & \progressbar{0.2}{1.4}{3.8}{3.6}{1}{\textbf{16\%}}{\textbf{46\%}} \\
        Recency of project releases   & \progressbar{0.5}{1.4}{3.6}{3.7}{0.8}{\textbf{18\%}}{\textbf{46\%}} \\
        Frequency of merging PRs from new contributors  & \progressbar{0.2}{2.2}{3.8}{2.8}{1}{\textbf{24\%}}{\textbf{38\%}} \\
        Having contributors from various countries  & \progressbar{3.5}{1.8}{2.7}{1.4}{0.6} {\textbf{53\%}}{\textbf{20\%}} \\
        Gender diversity & \progressbar{4.4}{1.5}{2.4}{1.3}{0.4}{\textbf{59\%}}{\textbf{17\%}} \\
        Having contributors from their own country              & \progressbar{3.5}{1.9}{3}{1}{0.6}{\textbf{54\%}}{\textbf{15\%}} \\

        \bottomrule
    \end{tabular}

    \vspace{0.5em}
    {\footnotesize
\begin{tabular}{l l l l l l l l l l}
  \cellcolor{NotAtAll} & Not at all important&
  \cellcolor{Slightly} & Slightly important&
  \cellcolor{Moderately} & Moderately important&
  \cellcolor{Very} & Very important&
  \cellcolor{Extremely} & Extremely important
\end{tabular}
}

\end{table*}

\begin{tcolorbox}[arc=1mm,width=1.0\columnwidth,
                  top=1mm,left=1mm,  right=1mm, bottom=1mm,
                  boxrule=.75pt]
 \textbf{RQ 1.1 Summary:} Practitioners prioritized fast, clear, and respectful communication when selecting OSS projects. Team dynamics like consistency of receiving responses (76\%), clarity and simplicity of replies (76\%), and tone of communication (73\%) were highly rated as “Very” or “Extremely” important. Responsiveness indicators (e.g., PR review and issue resolution times) and active member participation were also highly valued. Diversity-related team dynamics such as gender and regional mix were rated lower, though some respondents still emphasised inclusion in open-ended responses. 
\end{tcolorbox}

\subsubsection{RQ 1.2: What additional team dynamics do practitioners identify as important beyond those analyzed in RQ 1.1?}

The analysis of practitioners' additional recommended team dynamics beyond those in Section \ref{team_dynamics_1_1_1} has uncovered 8 more team dynamics, detailed below. We use the notation $\times$\emph{\textbf{i}} to indicate that \emph{\textbf{i}} times it was mentioned by survey participants.



\textbf{\textit {Willingness to collaborate} ($\times$35)} This team dynamic emerged as the most pronounced, with participants valuing friendly, accountable team members and their willingness to help and collaborate with one another.
\begin{quoting}
   ``\textit{How willing people are to offer help. If people aren't willing to help, then you can never get started.}'' (Developer)
\end{quoting}
\begin{quoting}
   ``\textit{I appreciate people's humility...}'' (Data Scientist)
\end{quoting}

Moreover, the responses expressed appreciation for having active and dedicated members.
\begin{quoting}
   ``\textit{I would rather join a project with a dedicated team that feels more like colleagues and less like a friend group, but also isn't obsessive or overly strict about it}'' (Developer)
\end{quoting}

\textbf{\textit{ Onboarding resources and assistance} ($\times$16)} Participants valued the projects that welcome new members, provide setup instructions and offer onboarding support and clear documentation.
\begin{quoting}
   ``\textit{Projects that invest in strong first issues, clear contribution guidelines, and setup instructions demonstrate their dedication to inclusion and accessibility. If I can simply get up and going, I'll be driven to help. Poor onboarding is frequently a red indicator for larger systemic issues.}'' (Developer)
\end{quoting}

\textbf{\textit{Inclusive and transparent decision-making} ($\times$15)} Participants emphasized the importance of a decision-making process that is transparent, inclusive, and collaborative.
\begin{quoting}
   ``\textit{Projects that have a clear, transparent, and collaborative decision-making process are more sustainable and inclusive.}'' (Developer)
\end{quoting}
 They further emphasized that this process will enhance their confidence and encourage greater participation. 
\begin{quoting}
   ``\textit{One important team dynamic to consider is the project's approach to collaboration and decision making because a transparent and inclusive process can make contributors feel valued and encourage greater participation.}'' (Technical Consultant)
\end{quoting}

\textbf{\textit{Providing mentorship and growth opportunities ($\times$11)}} Participants appreciated the communities that offer mentorship, opportunities to build their reputation, and chances for personal growth.
\begin{quoting}
   ``\textit{Highlighting contributors (to create incentives for contributions), structure (technical boards, decision boards etc.), aim to gain reputation in CNCF and other communities.}'' (Technical Product Manager)
\end{quoting}

\textbf{\textit{Providing constructive feedback ($\times$11)}} Participants valued the constructive and respectful feedback given by the project team.
\begin{quoting}
   ``\textit{Constructive feedback helps in good learning and growth}'' (Data Engineer)
\end{quoting}

\textbf{\textit{Effective communication channels and language ($\times$6)}} Having a good communication channel and skills were highlighted as important in the responses.
\begin{quoting}
   ``\textit{Something that can influence me to join an OSS project is if the team has a good communication channel/skills}'' (Developer)
\end{quoting}

Further, they noted that the primary language of communication significantly influences their decision to join an OSS project.
\begin{quoting}
   ``\textit{The primary language of communication. Even with AI assistance nowadays, communicating in languages other than English or one's native tongue can still be quite challenging.}'' (Data Scientist)
\end{quoting}

\textit{\textbf{Having clear and defined roles ($\times$5)}} Participants valued having well-defined roles within the team.
\begin{quoting}
   ``\textit{A well-organised team that clearly defines roles... signals maturity and sustainability.}'' (Developer)
\end{quoting}

\textit{\textbf{Miscellaneous} ($\times$7)} The conflict resolution mechanism was considered an important factor when selecting a project to join. This was mentioned $\times$4 in the responses.
\begin{quoting}
   ``\textit{I would also consider how the team handles conflict resolution and whether there’s a culture of knowledge sharing and mentorship. }'' (Developer)
\end{quoting}

Additionally, participants valued having personally known or contributors that they have previously collaborated with. This was mentioned $\times$3 in the responses.
\begin{quoting}
   ``\textit{Level of knowing the main devs, it would help in getting involved }'' (Developer)
\end{quoting}

\begin{tcolorbox}[arc=1mm,width=1.0\columnwidth,
                  top=1mm,left=1mm,  right=1mm, bottom=1mm,
                  boxrule=.75pt]
 \textbf{RQ 1.2 Summary:} Our study found eight additional team dynamics that practitioners consider important when selecting OSS projects. The most frequent were willingness to collaborate (×35), onboarding resources and assistance (×16), and inclusive and transparent decision-making (×15). Others included mentorship and growth opportunities (×11), constructive feedback (×11), effective communication channels and language (×6), and clear team roles (×5). These additional team dynamics highlight the social and procedural qualities practitioners associate with healthy OSS communities.
\end{tcolorbox}

\subsection{RQ 2: How do OSS practitioners’ motivations influence team dynamics when deciding to join OSS projects?}

We examined how OSS practitioners’ motivations influenced their preferences for team dynamics. Table \ref{pval_motivations_team_dynamics} summarizes the Kruskal–Wallis p-values. Only associations with statistically significant differences (p $\leq 0.05$ after Bonferroni correction) were further analyzed; p > 0.05 indicates no significant variation in how motivational groups rated the factor’s importance.

\begin{table*}[]
\caption{Kruskal Wallis P values for Team Dynamics and Motivations Analysis}
\label{pval_motivations_team_dynamics}
\resizebox{\textwidth}{!}{%
\begin{tabular}{llllllll}
\hline
 Team Dynamics & Learning & Helping & Enjoyment & Career & Networking & Reputation & Incentives \\ \hline
Having active members contributing regularly & \textbf{0.028} & \textbf{0.002} & \textbf{0.276} & 0.056 & \textbf{\textless{}.001} & \textbf{0.003} & \textbf{\textless{}.001} \\
Gender diversity & 0.091 & 0.08 & 0.950 & \textbf{0.005} & \textbf{0.005} & \textbf{\textless{}.001} & 0.052 \\
Having contributors from various countries & 0.058 & \textbf{0.017} & 0.585 & \textbf{0.031} & \textbf{\textless{}.001} & \textbf{\textless{}.001} & \textbf{0.012} \\
Having contributors from their own country & 0.157 & \textbf{0.035} & 0.415 & \textbf{0.04} & \textbf{0.002} & \textbf{\textless{}.001} & \textbf{0.014} \\
Consistency of receiving responses to questions & \textbf{\textless{}.001} & \textbf{0.002} & \textbf{0.039} & \textbf{\textless{}.001} & \textbf{0.005} & \textbf{0.03} & 0.245 \\
Typical response time to questions or inquiries & \textbf{\textless{}.001} & 0.067 & 0.09 & \textbf{\textless{}.001} & \textbf{0.018} & 0.15 & \textbf{0.045} \\
Clarity and simplicity of replies to questions & \textbf{0.001} & \textbf{\textless{}.001} & \textbf{0.043} & \textbf{0.015} & \textbf{0.007} & \textbf{0.007} & 0.318 \\
Tone and manner of communication & \textbf{\textless{}.001} & \textbf{0.007} & 0.753 & 0.121 & \textbf{\textless{}.001} & \textbf{\textless{}.001} & 0.199 \\
Frequency of merging PRs from new contributors & \textbf{\textless{}.001} & \textbf{\textless{}.001} & 0.256 & \textbf{0.002} & \textbf{\textless{}.001} & \textbf{\textless{}.001} & \textbf{0.01} \\
Whether PRs generally receive a clear response & \textbf{\textless{}.001} & 0.697 & 0.291 & \textbf{0.04} & 0.434 & 0.383 & 0.806 \\
Typical time taken to review or respond to PRs & \textbf{\textless{}.001} & \textbf{0.008} & 0.793 & \textbf{0.006} & \textbf{0.021} & \textbf{0.029} & 0.165 \\
Time taken to resolve reported issues & \textbf{0.003} & 0.058 & 0.219 & \textbf{\textless{}.001} & \textbf{0.016} & \textbf{0.028} & \textbf{0.013} \\
Availability of beginner-friendly issues & \textbf{\textless{}.001} & \textbf{0.009} & 0.525 & \textbf{0.007} & \textbf{\textless{}.001} & \textbf{0.041} & 0.12 \\
Use of descriptive labels on issues & \textbf{\textless{}.001} & \textbf{0.014} & \textbf{0.047} & \textbf{0.002} & \textbf{0.003} & \textbf{0.013} & 0.334 \\
Frequency of recent code commits & \textbf{\textless{}.001} & \textbf{0.003} & \textbf{0.039} & \textbf{0.009} & \textbf{0.041} & \textbf{0.023} & 0.087 \\
Recency of project releases & \textbf{\textless{}.001} & \textbf{\textless{}.001} & \textbf{0.007} & \textbf{0.005} & \textbf{0.003} & \textbf{0.044} & 0.123 \\
Level of activity in issue or pull request discussions & \textbf{\textless{}.001} & \textbf{\textless{}.001} & \textbf{0.003} & \textbf{0.003} & \textbf{0.006} & \textbf{0.046} & 0.054 \\ \hline
\end{tabular}%
}
\end{table*}

\subsubsection{Learning Motivation}

Participants with higher (rated “Very” or “Extremely”) learning motivation placed greater importance on having active members contributing regularly, consistency of receiving responses, typical response time to questions or inquiries, the clarity and simplicity of replies, and the tone and manner of communication. In terms of responsiveness and collaboration, frequency of merging PRs, whether PRs generally receive a clear response, the time taken to review or respond to PRs, time taken to resolve reported issues, avilability of beginner-friendly issues, use of descriptive labels, the frequency of recent code commits, recency of project releases,  and the level of activity in issue or pull request discussions were deemed highly important.

\subsubsection{Helping Motivation}

Participants with higher helping motivation rated having active members contributing regularly, having contributors from various countries, consistency of receiving responses, the clarity and simplicity of replies, and the tone and manner of communication as highly important. Moreover, the time taken to review or respond to PRs, the frequency of merging PRs, the availability of beginner-friendly issues, the use of descriptive labels, the frequency of recent code commits, the recency of project releases and the level of activity in issue or pull request discussions were also rated as highly important for them when selecting an OSS project to join. Although having contributors from their own country initially showed a statistically significant Kruskal–Wallis p-value (0.035) among motivational groups, it did not remain significant after applying the Bonferroni correction.

\subsubsection{Enjoyment Motivation}

Those who have a higher enjoyment motivation placed greater importance on having active members contributing regularly, the clarity and simplicity of replies, the frequency of recent code commits, the recency of project releases and the level of activity in issue or pull request discussions. Initially, consistency of receiving responses to questions and use of descriptive labels on issues showed statistically significant Kruskal–Wallis p-values among motivational groups; however, this significance did not hold after applying the Bonferroni correction.

\subsubsection{Career Motivation}

Participants with higher career motivation rated having contributors from their own country, consistency of receiving responses, and typical response time to questions or inquiries as important. In terms of responsiveness and collaboration, frequency of merging PRs, clarity and simplicity of replies to questions,  receiving clear responses to PRs, the time taken to review or respond to PRs, time to resolve reported issues, availability of beginner-friendly issues, use of descriptive labels, the frequency of recent code commits, recency of project releases and the level of activity in issue or pull request discussions were rated as highly important. Although gender diversity and having contributors from various countries initially showed statistically significant Kruskal–Wallis p-values among motivational groups, they did not remain significant after applying the Bonferroni correction.

\subsubsection{Networking Motivation}

Those who have a higher networking motivation placed greater importance on having active members contributing regularly, gender diversity, having contributors from their own country, and having contributors from various countries. Moreover, they consistently rated communication-related factors, the consistency of receiving responses, typical response time to questions or inquiries, the clarity and simplicity of replies, tone and manner of communication as highly important. The frequency of merging PRs, the time taken to review or respond to PRs, the time to resolve reported issues, the availability of beginner-friendly issues, the use of descriptive labels, the recency of project releases and the level of activity in issue or pull request discussions were also considered highly important. Although the frequency of recent code commits initially showed a statistically significant Kruskal–Wallis p-value (0.041) among motivational groups, it did not remain significant after applying the Bonferroni correction.

\subsubsection{Reputation Motivation}

Those who have a higher reputation motivation placed greater importance on having active members contributing regularly, gender diversity, having contributors from various countries, and having contributors from their own country. Moreover, clarity and simplicity of replies, tone and manner of communication, the frequency of merging PRs, the time to resolve reported issues, the frequency of recent code commits, the availability of beginner-friendly issues, the use of descriptive labels and the level of activity in issue or pull request discussions were also considered highly important by them. Although consistency of receiving responses to questions, typical time taken to review or respond to PRs, and recency of project releases initially showed statistically significant Kruskal–Wallis p-values among motivational groups, they did not remain significant after applying the Bonferroni correction.

\subsubsection{Incentives Motivation}

Those with higher incentives motivation placed greater importance on having active members contributing regularly, having contributors from various countries, and having contributors from their own country. The typical response time to questions or inquiries, the time to resolve reported issues, and the frequency of merging PRs were also deemed important.

\begin{tcolorbox}[arc=1mm,width=1.0\columnwidth,
                  top=1mm,left=1mm,  right=1mm, bottom=1mm,
                  boxrule=.75pt]
 \textbf{RQ 2 Summary:} Motivations shape how practitioners prioritise different team dynamics when choosing which OSS projects to join. Practitioners motivated by \textit{learning} or \textit{helping} preferred projects with supportive communication and collaborative team environments. Practitioners driven by \textit{career} preferred diverse communities which supports positive communication and collaboration.
 Those motivated by \textit{enjoyment} were linked to communities that demonstrated active collaboration, clear structure, and beginner-friendly practices. \textit{Networking} and \textit{reputation} driven contributors were linked to inclusive teams that encouraged diverse participation. Finally, contributors motivated by \textit{incentives} tended to select communities that maintained regional diversity, promptly replied to inquiries, merged new pull requests frequently, and resolved issues efficiently. These differences indicate that motivation-aware weighting could enhance the personalisation of OSS project recommendations.

\end{tcolorbox}

\section{Discussion} \label{discussion}

\subsection{Humans and AIs to co-maintain OSS projects}

Our RQ 1.1 findings reveal that OSS practitioners primarily focus on communication-related team dynamics, notably consistency of receiving responses (76\%), clarity and simplicity of replies (76\%), and tone and manner of communication (73\%). Practitioners also valued constructive feedback, mentorship, and transparency in decision-making as indicators of healthy, sustainable collaboration. These preferences emphasise healthy communication and collaboration as central social cues shaping practitioners’ willingness to join or remain in projects. However, maintaining such an environment at scale remains challenging, particularly as OSS teams expand globally and rely on asynchronous communication \cite{nirmani2025systematic}. Prior studies have raised concerns that fully automating communication processes with AI can erode empathy, trust, and the authenticity of human interaction within developer communities \cite{wang2024investigating, cheng2024would}. This creates an opportunity for AI-mediated collaboration systems that can enhance, rather than replace, human maintainers.

With the current advancements in AI, recent studies reveal how AI systems can actively co-maintain software projects alongside humans. For instance, Bots now automate routine pull request tasks, saving time and improving coordination \cite{wessel2022bots}. A recent study reveals that engineers perceive Large Language Models (LLM) generated code reviews as more polite and emotionally consistent, thereby reducing the need for emotional regulation and mitigating interpersonal friction during code reviews. However, it comes at the cost of a higher cognitive load when interpreting detailed feedback \cite{alami2025human}. A similar study revealed that the LLM-based code review tools improve bug detection and encourage better coding practices. Still, they also slow down pull request completion and sometimes produce incorrect or irrelevant feedback \cite{cihan2025automated}. They may overlook valuable human aspects, including knowledge sharing, shared code ownership, and team awareness. Therefore, existing studies suggest that AI should support human code reviewers rather than replace them, enhancing both code quality and collaboration \cite{heander2025support}.

Moreover, another area involves automated bug fixing and code refactoring in OSS projects. A recent study introduced a repair bot that continuously learns from ongoing code changes to develop effective bug-fixing strategies \cite{baudry2021software}. However, such systems have drawbacks, as maintainers often feel overwhelmed by the frequency of bot activities and the volume of notifications \cite{baudry2021software}. Documentation plays a critical role in OSS, and AI is now assisting maintainers in producing and managing documentation. A recent study addresses the lack of multilingual documentation in OSS projects \cite{kayode2025bridging}. It demonstrates that LLMs can assist in creating or updating translations, thereby making projects more accessible worldwide. Although LLMs are not yet completely reliable, they can provide useful first drafts and enhance translation quality when they have access to previous versions \cite{kayode2025bridging}.

\textbf{\underline{Implication:}} Future OSS ecosystems could adopt human-AI co-maintenance frameworks in which AI assistants act as responsiveness amplifiers that monitor PRs, detect stalled discussions, and prompt contributors with empathetic, contextually relevant reminders. Such systems could integrate emotion-aware communication models to maintain the respectful tone practitioners value most \cite{cheriyan2021towards}. For example, emotion-aware AI assistants could alert maintainers when discussion tone drifts from constructive norms or suggest phrasing that preserves psychological safety \cite{imran2025understanding}.

\subsection{Inclusive communication to promote diversity}

Diversity and inclusion in OSS projects promote valuable perspectives, experiences, and skills that come from contributors with varied backgrounds \cite{feng2023state}. It also promotes access for people from underrepresented groups to the learning and career opportunities that participation in OSS can provide, making it both a practical necessity and a broader social responsibility \cite{feng2023state}. Although diversity-related factors such as gender representation or regional mix received relatively lower ratings in our quantitative results (RQ 1.1), several participants who rated gender and regional diversity as very or extremely important in the closed-ended questions also re-emphasised the importance of diversity and inclusion in their open-ended responses. In practice, contributors appear to value diversity not as a standalone goal, but as something that thrives when teams foster respectful, open, and empathetic communication.

A recent study revealed that male and female developers showed similar productivity in OSS \cite{bosu2019diversity}. However, in some projects, women faced bias, as evidenced by lower code acceptance rates and slower review feedback \cite{bosu2019diversity}. As a remedy, they suggest building dedicated communities or platforms for women in OSS that can help them share experiences, build confidence, and feel more supported, ultimately reducing silent dropout and improving diversity. Another study revealed that non-inclusive communication, toxic culture, and stereotyping often hinder participation from underrepresented groups \cite{feng2023state}. The authors recommend adopting multiple communication channels and local language groups to help bridge cultural and language gaps and make contributors feel more included.

A literature survey on women’s participation in OSS found that the main challenges they face are non-inclusive communication, harassment, and toxic project cultures \cite{trinkenreich2022women}. Hostile or unwelcoming interactions often drive women away from contributing. To address this, the authors recommend promoting inclusive language and creating and enforcing a clear Code of Conduct that defines acceptable behaviour, prohibits harassment, and includes transparent mechanisms for reporting and enforcing consequences. Another study revealed that many OSS communities lack enforced codes of conduct, making it hard for newcomers to assess community quality \cite{prana2021including}. They further revealed that promoting such codes and introducing social metrics could improve transparency. Additionally, creating ways to highlight contributors from underrepresented groups and offering better translation tools can help increase visibility and participation among minority developers \cite{prana2021including}. These findings align with the growing evidence that communication and diversity are interdependent, rather than competing forces.

\textbf{\underline{Implication:}} Diversity initiatives in OSS communities should be complemented by efforts to strengthen inclusive communication practices. Open, empathetic dialogues and a clear code of conduct allow diversity to translate into collaboration, creativity, and long-term community health.

\subsection{Transparent, collaborative decision-making drives perceived fairness}

Our open-ended responses (RQ 1.2) revealed that contributors highly value transparent, inclusive, and collaborative decision-making when choosing OSS projects. Practitioners described such processes as signals of fairness, maturity, and sustainability, which increase their confidence and willingness to participate. For instance, they mentioned that, “\textit{Inclusive decision-making ensures all voices matter, making me more confident and willing to contribute.}” (Developer)
They further highlighted the importance of conflict resolution mechanisms and the need to distribute authority more equitably among maintainers and contributors (RQ 1.2).

Prior research supports this finding. A recent study found that contributors in projects led by a single “benevolent dictator” often faced frustration due to limited collaborative decision-making \cite{geiger2021labor}. In contrast, transparent and participatory governance increased contributors’ confidence, engagement, and project sustainability. They recommend that OSS projects document their decision-making processes publicly to ensure that contributors can understand how and why decisions are made. Adopting consensus-based governance methods, such as community voting or structured discussions, fosters collaboration and shared ownership among contributors. Additionally, implementing rotating or shared leadership roles helps balance responsibilities and supports inclusive, scalable project management \cite{geiger2021labor}. 

Another study found that imbalanced power structures where a few maintainers control decision-making or access can discourage participation and lead to contributor dropout \cite{farias2021power}. Particularly, newcomers often face barriers when their contributions are rejected due to hierarchical structures and limited familiarity with core members. To address these challenges, the authors recommend making power relations explicit within projects, allowing community members to identify and mitigate coercive or exclusionary dynamics. They suggest fostering reward and referent-based power, where recognition, learning, and mentorship motivate participation, while minimizing coercive or legalistic power that centralizes control \cite{farias2021power}. Similarly, another work highlighted that clear conflict resolution mechanisms (e.g. well-defined codes of conduct and dispute protocols) are vital for community health \cite{cobos2025bot}. These mechanisms ensure accountability and help resolve disagreements fairly and transparently, which in turn boosts contributors’ trust that the project is being run in an impartial, sustainable manner.

\textbf{\underline{Implication:}} OSS communities should promote transparent, participatory governance by sharing decisions openly, involving contributors in discussions, and rotating leadership roles. These practices build trust, fairness, and long-term project sustainability.

\subsection{Using Team Dynamics to Improve Automated OSS Project Recommendations}

Our findings provide a strong foundation for integrating human-centric team dynamics into future automated project recommendation systems. Current OSS project recommendation systems mainly rely on technical factors; however, integrating human factors can enhance the relevance and sustainability of recommendations \cite{nirmani2025systematic, abbasi2021relevant,phatangare2024codecompass, shen2024multi}. From RQ1.1, we found that practitioners consistently prioritize communication, collaboration, and responsiveness signals when selecting an OSS project to contribute to. They rated team dynamics, e.g., gender and regional diversity representation lower. From RQ1.2, the open-ended data provided additional insight, highlighting onboarding resources and assistance, transparent decision-making, constructive feedback, mentorship, etc., as key indicators of healthy teams. Together, these results establish a stable “social baseline” that most practitioners seek before considering joining a project. 

Moreover, RQ2 shows how motivations modulate this baseline. The preferences for team dynamics varied according to contributors' motivations for joining OSS. The relationship between motivation and team dynamics captures how contributors interpret and weigh social signals differently based on what drives them to participate. For instance, those motivated by networking or reputation value diversity and inclusiveness within the community. Conversely, those driven by enjoyment appreciate beginner-friendly, structured, and active collaboration practices. Studying this correlation, therefore, offers a deeper understanding of contributor-project fit, revealing how motivation reshapes the perceived importance of team behaviours. Hence, recommendations can be personalized via motivation-aware weighting based on our RQ2 results. These findings can be used to compute a repository social signal vector to measure the project's health and re-rank recommendations.

\textbf{\underline{Implication:}} Automated OSS project recommenders should move beyond technical fit and re-rank candidates using a motivation-aware weighting of empirically important team-dynamics signals that improve both relevance and long-term contributor–project fit.

\section{Threats to validity} \label{threats}
\textbf{Internal Validity:} Several factors may have adversely affected the accuracy of our data collection and analysis. Some participants might not have had the required experience or expertise for our study. To address this, we included a screening test before the survey started. Participants’ perceptions of team dynamics may have been influenced by recall bias \cite{tayeb2024investigating,gerosa2021shifting} (which occurs when participants fail to remember past details accurately) or personal interpretations. To minimize this, we clearly defined each construct in the questionnaire and included examples to support consistent understanding. Moreover, a pilot study was conducted with software practitioners to refine question wording and improve clarity before the main survey. Another possible threat is social desirability bias \cite{tayeb2024investigating}, where participants might provide answers that reflect favorable team practices rather than their genuine opinions. To mitigate this, the survey was anonymous, and no identifiable information was collected. Although a non-parametric test was used, the relatively small sample size (198 responses) may have limited our ability to identify all significant correlations in the analysis. Finally, while we selected the most relevant team dynamic factors based on prior literature, our list may not fully encompass all possible aspects that influence collaboration in OSS teams.

\textbf{Construct Validity:} One possible limitation of our survey was the risk of missing relevant options, which could lead to underrepresentation or bias in the results. To mitigate this, we provided an “Other” option in applicable questions. However, the findings indicate that most participants selected “Other” as their least preferred choice, suggesting that the survey captured the most relevant options. Another potential limitation was the subjectivity in interpreting responses during thematic analysis. To minimize this, three annotators independently reviewed and categorized the responses, and any disagreements were resolved through discussion.

\textbf{External Validity:} Our findings are based on responses from 198 OSS practitioners and may not generalize to the entire OSS contributor population. Overrepresentation of particular demographics or regions could limit the applicability of our results to other OSS contexts. Moreover, our survey focused on English-speaking participants, which might bias responses toward projects and communication styles prevalent in Western OSS ecosystems. Future studies should replicate this work with a more diverse participant base and explore other communication cultures and project governance models to strengthen generalizability.

\section{Conclusion and Future Works} \label{conclusion}
This study provides empirical insights into how social and behavioral aspects of collaboration influence OSS practitioners’ project selection. Contributors consistently prioritized fast and constructive responses and valued inclusive communication, while diversity-related aspects such as gender and regional representation were rated less critical overall. However, the results also showed that the perceived importance of team dynamics differs across motivations. For instance, contributors with strong reputation-gaining motivation especially valued projects that showed high levels of activity in issue and pull request discussions. Integrating such human-centric insights into OSS recommendation models would enable systems that not only match skills and technical fit but also promote sustainable, supportive communities.

Future work could advance this study in several meaningful directions.
First, our results showed that contributors consistently prioritize responsiveness and communication quality; hence, future research could quantify these social signals directly from repository activity (e.g., response times or thread length) to build measurable indicators of team health. While this study captured perceived team qualities, longitudinal analyses could examine how these perceptions evolve after contributors join a project, revealing whether early impressions of communication and responsiveness predict long-term retention. Finally, expanding this investigation to diverse collaboration contexts such as corporate-backed OSS or volunteer teams could uncover whether the same team dynamics hold across varying governance and incentive structures.

\bibliographystyle{ACM-Reference-Format}
\bibliography{ref}

@article{nirmani2025systematic,
  title={A systematic literature review on task recommendation systems for crowdsourced software engineering},
  author={Nirmani, Shashiwadana and Shahin, Mojtaba and Khalajzadeh, Hourieh and Liu, Xiao},
  journal={Information and Software Technology},
  pages={107753},
  year={2025},
  publisher={Elsevier}
}

@inproceedings{alami2025human,
  title={Human and Machine: How Software Engineers Perceive and Engage with AI-Assisted Code Reviews Compared to Their Peers},
  author={Alami, Adam and Ernst, Neil},
  booktitle={2025 IEEE/ACM 18th International Conference on Cooperative and Human Aspects of Software Engineering (CHASE)},
  pages={63--74},
  year={2025},
  organization={IEEE}
}

@inproceedings{cihan2025automated,
  title={Automated code review in practice},
  author={Cihan, Umut and Haratian, Vahid and {\.I}{\c{c}}{\"o}z, Arda and G{\"u}l, Mert Kaan and Devran, {\"O}mercan and Bayendur, Emircan Furkan and U{\c{c}}ar, Baykal Mehmet and T{\"u}z{\"u}n, Eray},
  booktitle={2025 IEEE/ACM 47th International Conference on Software Engineering: Software Engineering in Practice (ICSE-SEIP)},
  pages={425--436},
  year={2025},
  organization={IEEE}
}

@inproceedings{heander2025support,
  title={Support, not automation: towards AI-supported code review for code quality and beyond},
  author={Heander, Lo and S{\"o}derberg, Emma and Rydenf{\"a}lt, Christofer},
  booktitle={Proceedings of the 33rd ACM International Conference on the Foundations of Software Engineering},
  pages={591--595},
  year={2025}
}

@inproceedings{wessel2022bots,
  title={Bots for pull requests: The good, the bad, and the promising},
  author={Wessel, Mairieli and Abdellatif, Ahmad and Wiese, Igor and Conte, Tayana and Shihab, Emad and Gerosa, Marco A and Steinmacher, Igor},
  booktitle={Proceedings of the 44th International Conference on Software Engineering},
  pages={274--286},
  year={2022}
}

@article{baudry2021software,
  title={A software-repair robot based on continual learning},
  author={Baudry, Benoit and Chen, Zimin and Etemadi, Khashayar and Fu, Han and Ginelli, Davide and Kommrusch, Steve and Martinez, Matias and Monperrus, Martin and Ron, Javier and Ye, He and others},
  journal={IEEE Software},
  volume={38},
  number={4},
  pages={28--35},
  year={2021},
  publisher={IEEE}
}

@article{kayode2025bridging,
  title={Bridging Language Gaps in Open-Source Documentation with Large-Language-Model Translation},
  author={Kayode Adejumo, Elijah and Johnson, Brittany and Guizani, Mariam},
  journal={arXiv e-prints},
  pages={arXiv--2508},
  year={2025}
}

@inproceedings{cheriyan2021towards,
  title={Towards offensive language detection and reduction in four software engineering communities},
  author={Cheriyan, Jithin and Savarimuthu, Bastin Tony Roy and Cranefield, Stephen},
  booktitle={Proceedings of the 25th International Conference on Evaluation and Assessment in Software Engineering},
  pages={254--259},
  year={2021}
}

@article{imran2025understanding,
  title={Understanding and predicting derailment in toxic conversations on GitHub},
  author={Imran, Mia Mohammad and Zita, Robert and Copeland, Rebekah and Chatterjee, Preetha and Rahman, Rahat Rizvi and Damevski, Kostadin},
  journal={arXiv preprint arXiv:2503.02191},
  year={2025}
}

@article{feng2023state,
  title={The state of diversity and inclusion in apache: A pulse check},
  author={Feng, Zixuan and Guizani, Mariam and Gerosa, Marco A and Sarma, Anita},
  journal={arXiv preprint arXiv:2303.16344},
  year={2023}
}

@inproceedings{bosu2019diversity,
  title={Diversity and inclusion in open source software (OSS) projects: Where do we stand?},
  author={Bosu, Amiangshu and Sultana, Kazi Zakia},
  booktitle={2019 ACM/IEEE International Symposium on Empirical Software Engineering and Measurement (ESEM)},
  pages={1--11},
  year={2019},
  organization={IEEE}
}

@article{trinkenreich2022women,
  title={Women’s participation in open source software: A survey of the literature},
  author={Trinkenreich, Bianca and Wiese, Igor and Sarma, Anita and Gerosa, Marco and Steinmacher, Igor},
  journal={ACM Transactions on Software Engineering and Methodology (TOSEM)},
  volume={31},
  number={4},
  pages={1--37},
  year={2022},
  publisher={ACM New York, NY}
}

@article{prana2021including,
  title={Including everyone, everywhere: Understanding opportunities and challenges of geographic gender-inclusion in oss},
  author={Prana, Gede Artha Azriadi and Ford, Denae and Rastogi, Ayushi and Lo, David and Purandare, Rahul and Nagappan, Nachiappan},
  journal={IEEE Transactions on Software Engineering},
  volume={48},
  number={9},
  pages={3394--3409},
  year={2021},
  publisher={IEEE}
}

@article{geiger2021labor,
  title={The labor of maintaining and scaling free and open-source software projects},
  author={Geiger, R Stuart and Howard, Dorothy and Irani, Lilly},
  journal={Proceedings of the ACM on human-computer interaction},
  volume={5},
  number={CSCW1},
  pages={1--28},
  year={2021},
  publisher={ACM New York, NY, USA}
}

@inproceedings{farias2021power,
  title={Power relations within an open source software ecosystem},
  author={Farias, Victor and Wiese, Igor and Santos, Rodrigo},
  booktitle={International Conference on Software Business},
  pages={187--193},
  year={2021},
  organization={Springer}
}

@inproceedings{cobos2025bot,
  title={A Bot-Based Approach to Manage Codes of Conduct in Open-Source Projects},
  author={Cobos, Sergio and Izquierdo, Javier Luis C{\'a}novas},
  booktitle={2025 IEEE/ACM 47th International Conference on Software Engineering: Software Engineering in Society (ICSE-SEIS)},
  pages={59--69},
  year={2025},
  organization={IEEE}
}

@article{abbasi2021relevant,
  title={Relevant projectrecommendation system using developer behavior and project features},
  author={Abbasi, Soohan and Ali, Zulfiqar and Kumar, Rajesh and Shaikh, Madiha and Maheshwari, Komal and Lal, Paras and Kumar, Sagar},
  journal={International Journal},
  volume={10},
  number={5},
  year={2021}
}

@inproceedings{phatangare2024codecompass,
  title={CodeCompass: NLP-driven navigation to optimal repositories},
  author={Phatangare, Sheetal and Matkar, Aakash and Jadhav, Akshay and Shaikh, Al Hussain and Bonde, Anish},
  booktitle={2024 4th International Conference on Pervasive Computing and Social Networking (ICPCSN)},
  pages={393--401},
  year={2024},
  organization={IEEE}
}

@article{shen2024multi,
  title={Multi-objective optimization and integrated indicator-driven two-stage project recommendation in time-dependent software ecosystem},
  author={Shen, Xin and Yao, Xiangjuan and Gong, Dunwei and Tu, Huijie},
  journal={Information and Software Technology},
  volume={170},
  pages={107433},
  year={2024},
  publisher={Elsevier}
}

@incollection{kitchenham2008personal,
  title={Personal opinion surveys},
  author={Kitchenham, Barbara A and Pfleeger, Shari L},
  booktitle={Guide to advanced empirical software engineering},
  pages={63--92},
  year={2008},
  publisher={Springer}
}

@article{hertel2003motivation,
  title={Motivation of software developers in Open Source projects: an Internet-based survey of contributors to the Linux kernel},
  author={Hertel, Guido and Niedner, Sven and Herrmann, Stefanie},
  journal={Research policy},
  volume={32},
  number={7},
  pages={1159--1177},
  year={2003},
  publisher={Elsevier}
}

@article{ke2009motivations,
  title={Motivations in open source software communities: The mediating role of effort intensity and goal commitment},
  author={Ke, Weiling and Zhang, Ping},
  journal={International Journal of Electronic Commerce},
  volume={13},
  number={4},
  pages={39--66},
  year={2009},
  publisher={Taylor \& Francis}
}

@article{von2012carrots,
  title={Carrots and rainbows: Motivation and social practice in open source software development},
  author={Von Krogh, Georg and Haefliger, Stefan and Spaeth, Sebastian and Wallin, Martin W},
  journal={MIS quarterly},
  pages={649--676},
  year={2012},
  publisher={JSTOR}
}

@article{lakhani2003hackers,
  title={Why hackers do what they do: Understanding motivation and effort in free/open source software projects},
  author={Lakhani, Karim R and Wolf, Robert G},
  journal={Open Source Software Projects (September 2003)},
  year={2003}
}

@article{wu2007empirical,
  title={An empirical analysis of open source software developers’ motivations and continuance intentions},
  author={Wu, Chorng-Guang and Gerlach, James H and Young, Clifford E},
  journal={Information \& Management},
  volume={44},
  number={3},
  pages={253--262},
  year={2007},
  publisher={Elsevier}
}

@article{roberts2006understanding,
  title={Understanding the motivations, participation, and performance of open source software developers: A longitudinal study of the Apache projects},
  author={Roberts, Jeffrey A and Hann, Il-Horn and Slaughter, Sandra A},
  journal={Management science},
  volume={52},
  number={7},
  pages={984--999},
  year={2006},
  publisher={INFORMS}
}

@inproceedings{danilova2021you,
  title={Do you really code? designing and evaluating screening questions for online surveys with programmers},
  author={Danilova, Anastasia and Naiakshina, Alena and Horstmann, Stefan and Smith, Matthew},
  booktitle={2021 IEEE/ACM 43rd International Conference on Software Engineering (ICSE)},
  pages={537--548},
  year={2021},
  organization={IEEE}
}

@article{mckight2010kruskal,
  title={Kruskal-wallis test},
  author={McKight, Patrick E and Najab, Julius},
  journal={The corsini encyclopedia of psychology},
  pages={1--1},
  year={2010},
  publisher={Wiley Online Library}
}

@article{mcknight2010mann,
  title={Mann-whitney U test},
  author={McKnight, Patrick E and Najab, Julius},
  journal={The Corsini encyclopedia of psychology},
  pages={1--1},
  year={2010},
  publisher={Wiley Online Library}
}

@inproceedings{aldndni2024understanding,
  title={Understanding the Impact of Branch Edit Features for the Automatic Prediction of Merge Conflict Resolutions},
  author={Aldndni, Waad and Servant, Francisco and Meng, Na},
  booktitle={Proceedings of the 32nd IEEE/ACM International Conference on Program Comprehension},
  pages={149--160},
  year={2024}
}

@inproceedings{meem2024exploring,
  title={Exploring experiences with automated program repair in practice},
  author={Meem, Fairuz Nawer and Smith, Justin and Johnson, Brittany},
  booktitle={Proceedings of the IEEE/ACM 46th International Conference on Software Engineering},
  pages={1--11},
  year={2024}
}

@article{ouyang2025empirical,
  title={An empirical study of the non-determinism of chatgpt in code generation},
  author={Ouyang, Shuyin and Zhang, Jie M and Harman, Mark and Wang, Meng},
  journal={ACM Transactions on Software Engineering and Methodology},
  volume={34},
  number={2},
  pages={1--28},
  year={2025},
  publisher={ACM New York, NY}
}

@article{weisstein2004bonferroni,
  title={Bonferroni correction},
  author={Weisstein, Eric W},
  journal={https://mathworld. wolfram. com/},
  year={2004},
  publisher={Wolfram Research, Inc.}
}

@online{Prolific,
  title        = {Prolific: Easily collect high-quality data from real people},
  howpublished = {\url{https://www.prolific.com/}},
  note         = {Accessed: 2025-10-10}
}

@article{reid2022software,
  title={Software engineering user study recruitment on prolific: An experience report},
  author={Reid, Brittany and Wagner, Markus and d'Amorim, Marcelo and Treude, Christoph},
  journal={arXiv preprint arXiv:2201.05348},
  year={2022}
}

@article{russo2024navigating,
  title={Navigating the complexity of generative ai adoption in software engineering},
  author={Russo, Daniel},
  journal={ACM Transactions on Software Engineering and Methodology},
  volume={33},
  number={5},
  pages={1--50},
  year={2024},
  publisher={ACM New York, NY}
}

@inproceedings{tahaei2023stuck,
  title={Stuck in the permissions with you: Developer \& end-user perspectives on app permissions \& their privacy ramifications},
  author={Tahaei, Mohammad and Abu-Salma, Ruba and Rashid, Awais},
  booktitle={Proceedings of the 2023 CHI Conference on Human Factors in Computing Systems},
  pages={1--24},
  year={2023}
}

@online{LinkedIn,
  title        = {LinkedIn},
  howpublished = {\url{https://www.linkedin.com/}},
  note         = {Accessed: 2025-10-10}
}

@inproceedings{yue2025discovering,
  title={Discovering Ideologies of the Open Source Software Movement},
  author={Yue, Yang and Wang, Yi and Redmiles, David},
  booktitle={2025 IEEE/ACM 47th International Conference on Software Engineering: New Ideas and Emerging Results (ICSE-NIER)},
  pages={21--25},
  year={2025},
  organization={IEEE}
}

@article{linaaker2025public,
  title={Public sector open source software projects-How is development organized?},
  author={Lin{\aa}ker, Johan and Lundell, Bj{\"o}rn and Servant, Francisco and Gamalielsson, Jonas and Muto, Sachiko and Robles, Gregorio},
  journal={Empirical Software Engineering},
  volume={30},
  number={3},
  pages={1--54},
  year={2025},
  publisher={Springer}
}

@article{steinmacher2019overcoming,
  title={Overcoming social barriers when contributing to open source software projects},
  author={Steinmacher, Igor and Gerosa, Marco and Conte, Tayana U and Redmiles, David F},
  journal={Computer Supported Cooperative Work (CSCW)},
  volume={28},
  number={1},
  pages={247--290},
  year={2019},
  publisher={Springer}
}

@article{iaffaldano2019developers,
  title={Why do developers take breaks from contributing to OSS projects? A preliminary analysis},
  author={Iaffaldano, Giuseppe and Steinmacher, Igor and Calefato, Fabio and Gerosa, Marco and Lanubile, Filippo},
  journal={arXiv preprint arXiv:1903.09528},
  year={2019}
}

@inproceedings{miller2022did,
  title={" Did you miss my comment or what?" understanding toxicity in open source discussions},
  author={Miller, Courtney and Cohen, Sophie and Klug, Daniel and Vasilescu, Bogdan and KaUstner, Christian},
  booktitle={Proceedings of the 44th international conference on software engineering},
  pages={710--722},
  year={2022}
}

@article{qiu2019signals,
  title={The signals that potential contributors look for when choosing open-source projects},
  author={Qiu, Huilian Sophie and Li, Yucen Lily and Padala, Susmita and Sarma, Anita and Vasilescu, Bogdan},
  journal={Proceedings of the ACM on Human-Computer Interaction},
  volume={3},
  number={CSCW},
  pages={1--29},
  year={2019},
  publisher={ACM New York, NY, USA}
}

@inproceedings{gerosa2021shifting,
  title={The shifting sands of motivation: Revisiting what drives contributors in open source},
  author={Gerosa, Marco and Wiese, Igor and Trinkenreich, Bianca and Link, Georg and Robles, Gregorio and Treude, Christoph and Steinmacher, Igor and Sarma, Anita},
  booktitle={2021 IEEE/ACM 43rd International Conference on Software Engineering (ICSE)},
  pages={1046--1058},
  year={2021},
  organization={IEEE}
}

@inproceedings{tayeb2024investigating,
  title={Investigating Developers' Preferences for Learning and Issue Resolution Resources in the ChatGPT Era},
  author={Tayeb, Ahmad and Alahmadi, Mohammad and Tajik, Elham and Haiduc, Sonia},
  booktitle={2024 IEEE International Conference on Software Maintenance and Evolution (ICSME)},
  pages={413--425},
  year={2024},
  organization={IEEE}
}

@inproceedings{steinmacher2014barriers,
  title={Barriers faced by newcomers to open source projects: a systematic review},
  author={Steinmacher, Igor and Silva, Marco Aur{\'e}lio Graciotto and Gerosa, Marco Aur{\'e}lio},
  booktitle={IFIP International Conference on Open Source Systems},
  pages={153--163},
  year={2014},
  organization={Springer}
}

@inproceedings{steinmacher2015social,
  title={Social barriers faced by newcomers placing their first contribution in open source software projects},
  author={Steinmacher, Igor and Conte, Tayana and Gerosa, Marco Aur{\'e}lio and Redmiles, David},
  booktitle={Proceedings of the 18th ACM conference on Computer supported cooperative work \& social computing},
  pages={1379--1392},
  year={2015}
}

@article{guizani2021long,
  title={The long road ahead: Ongoing challenges in contributing to large oss organizations and what to do},
  author={Guizani, Mariam and Chatterjee, Amreeta and Trinkenreich, Bianca and May, Mary Evelyn and Noa-Guevara, Geraldine J and Russell, Liam James and Cuevas Zambrano, Griselda G and Izquierdo-Cortazar, Daniel and Steinmacher, Igor and Gerosa, Marco A and others},
  journal={Proceedings of the ACM on Human-Computer Interaction},
  volume={5},
  number={CSCW2},
  pages={1--30},
  year={2021},
  publisher={ACM New York, NY, USA}
}

@inproceedings{gousios2016work,
  title={Work practices and challenges in pull-based development: The contributor's perspective},
  author={Gousios, Georgios and Storey, Margaret-Anne and Bacchelli, Alberto},
  booktitle={Proceedings of the 38th international conference on software engineering},
  pages={285--296},
  year={2016}
}

@inproceedings{jarczyk2014github,
  title={Github projects. quality analysis of open-source software},
  author={Jarczyk, Oskar and Gruszka, B{\l}a{\.z}ej and Jaroszewicz, Szymon and Bukowski, Leszek and Wierzbicki, Adam},
  booktitle={International Conference on Social Informatics},
  pages={80--94},
  year={2014},
  organization={Springer}
}

@inproceedings{ehsani2025analyzing,
  title={Analyzing Toxicity in Open Source Software Communications Using Psycholinguistics and Moral Foundations Theory},
  author={Ehsani, Ramtin and Rezapour, Rezvaneh and Chatterjee, Preetha},
  booktitle={2025 IEEE/ACM International Workshop on Natural Language-Based Software Engineering (NLBSE)},
  pages={1--8},
  year={2025},
  organization={IEEE}
}

@article{turzo2025first,
  title={From First Patch to Long-Term Contributor: Evaluating Onboarding Recommendations for OSS Newcomers},
  author={Turzo, Asif Kamal and Sultana, Sayma and Bosu, Amiangshu},
  journal={IEEE Transactions on Software Engineering},
  year={2025},
  publisher={IEEE}
}

@inproceedings{tsay2014influence,
  title={Influence of social and technical factors for evaluating contribution in GitHub},
  author={Tsay, Jason and Dabbish, Laura and Herbsleb, James},
  booktitle={Proceedings of the 36th international conference on Software engineering},
  pages={356--366},
  year={2014}
}

@article{terrell2017gender,
  title={Gender differences and bias in open source: Pull request acceptance of women versus men},
  author={Terrell, Josh and Kofink, Andrew and Middleton, Justin and Rainear, Clarissa and Murphy-Hill, Emerson and Parnin, Chris and Stallings, Jon},
  journal={PeerJ Computer Science},
  volume={3},
  pages={e111},
  year={2017},
  publisher={PeerJ Inc.}
}

@inproceedings{tan2020first,
  title={A first look at good first issues on GitHub},
  author={Tan, Xin and Zhou, Minghui and Sun, Zeyu},
  booktitle={Proceedings of the 28th ACM Joint Meeting on European Software Engineering Conference and Symposium on the Foundations of Software Engineering},
  pages={398--409},
  year={2020}
}

@article{liao2023graph,
  title={Graph convolutional network-based repository recommendation system},
  author={Liao, Zhifang and Cao, Shuyuan and Li, Bin and Liu, Shengzong and Zhang, Yan and Yu, Song},
  journal={Computer Modeling in Engineering \& Sciences},
  volume={137},
  number={1},
  pages={175--196},
  year={2023},
  publisher={Tech Science Press}
}

@inproceedings{sayce2022recommendation,
  title={Recommendation system for open source projects for minimizing abandonment},
  author={Sayce, Sarah and Ghosh, Krishnendu},
  booktitle={The International FLAIRS Conference Proceedings},
  volume={35},
  year={2022}
}

@inproceedings{ford2022reboc,
  title={Reboc: recommending bespoke open source software projects to contributors},
  author={Ford, Denae and Shrestha, Nischal and Zimmermann, Thomas},
  booktitle={2022 IEEE Symposium on Visual Languages and Human-Centric Computing (VL/HCC)},
  pages={1--5},
  year={2022},
  organization={IEEE}
}

@inproceedings{li2016task,
  title={Task recommendation with developer social network in software crowdsourcing},
  author={Li, Ning and Mo, Wenkai and Shen, Beijun},
  booktitle={2016 23rd Asia-Pacific Software Engineering Conference (APSEC)},
  pages={9--16},
  year={2016},
  organization={IEEE}
}

@inproceedings{santos2023tell,
  title={Tell me who are you talking to and i will tell you what issues need your skills},
  author={Santos, Fabio and Penney, Jacob and Pimentel, Jo{\~a}o Felipe and Wiese, Igor and Steinmacher, Igor and Gerosa, Marco A},
  booktitle={2023 IEEE/ACM 20th International Conference on Mining Software Repositories (MSR)},
  pages={611--623},
  year={2023},
  organization={IEEE}
}

@article{braun2006using,
  title={Using thematic analysis in psychology},
  author={Braun, Virginia and Clarke, Victoria},
  journal={Qualitative research in psychology},
  volume={3},
  number={2},
  pages={77--101},
  year={2006},
  publisher={Taylor \& Francis}
}

@article{smirnova2022makes,
  title={What makes the right OSS contributor tick? Treatments to motivate high-skilled developers},
  author={Smirnova, Inna and Reitzig, Markus and Alexy, Oliver},
  journal={Research Policy},
  volume={51},
  number={1},
  pages={104368},
  year={2022},
  publisher={Elsevier}
}

@article{sesari2025safe,
  title={Safe to Stay: Psychological Safety Sustains Participation in Pull-based Open Source Projects},
  author={Sesari, Emeralda and Sarro, Federica and Rastogi, Ayushi},
  journal={arXiv preprint arXiv:2504.17510},
  year={2025}
}

@inproceedings{wang2024investigating,
  title={Investigating and designing for trust in ai-powered code generation tools},
  author={Wang, Ruotong and Cheng, Ruijia and Ford, Denae and Zimmermann, Thomas},
  booktitle={Proceedings of the 2024 ACM Conference on Fairness, Accountability, and Transparency},
  pages={1475--1493},
  year={2024}
}

@article{cheng2024would,
  title={“It would work for me too”: How online communities shape software developers’ trust in AI-powered code generation tools},
  author={Cheng, Ruijia and Wang, Ruotong and Zimmermann, Thomas and Ford, Denae},
  journal={ACM Transactions on Interactive Intelligent Systems},
  volume={14},
  number={2},
  pages={1--39},
  year={2024},
  publisher={ACM New York, NY}
}

@dataset{anonymous_2025_beyondcode,
  author       = {Anonymous},
  title        = {Beyond Code: Empirical Insights into How Team Dynamics Influence OSS Project Selection},
  year         = {2025},
  version      = {1.0},
  publisher    = {Zenodo},
  doi          = {10.5281/zenodo.17420802},
  url          = {https://doi.org/10.5281/zenodo.17420802}
}

@inproceedings{ye2003toward,
  title={Toward an understanding of the motivation of open source software developers},
  author={Ye, Yunwen and Kishida, Kouichi},
  booktitle={25th International Conference on Software Engineering, 2003. Proceedings.},
  pages={419--429},
  year={2003},
  organization={IEEE}
}

@article{bitzer2007intrinsic,
  title={Intrinsic motivation in open source software development},
  author={Bitzer, J{\"u}rgen and Schrettl, Wolfram and Schr{\"o}der, Philipp JH},
  journal={Journal of comparative economics},
  volume={35},
  number={1},
  pages={160--169},
  year={2007},
  publisher={Elsevier}
}

@misc{hartman2011intrinsic,
  title={How do intrinsic and extrinsic motivation correlate with each other in open source software development?},
  author={Hartman, Kim},
  year={2011}
}

@article{li2012leadership,
  title={Leadership characteristics and developers’ motivation in open source software development},
  author={Li, Yan and Tan, Chuan-Hoo and Teo, Hock-Hai},
  journal={Information \& Management},
  volume={49},
  number={5},
  pages={257--267},
  year={2012},
  publisher={Elsevier}
}

@article{wynn2004leadership,
  title={Leadership and motivation in open source projects},
  author={Wynn Jr, Donald E},
  year={2004}
}
\end{document}